\documentclass[twocolumn,pre,a4paper,showpacs]{revtex4}
\begin{document} 
\title{Inclusive Spectra and Quantum Stochastic Processes}
\author{ A. Krzywicki\footnote{E-mail address: krz@th.u-psud.fr}}
\affiliation{Laboratoire de Physique Th\'eorique, B\^atiment 210,\\            
Universit\'e Paris-Sud,           
91405~Orsay, France}

\begin{abstract} 
We construct a simple, exactly solvable model
illustrating the conjecture that the evolution of the wave 
function of a system of very many interacting particles is likely 
to drive the one-particle spectrum towards a "thermal" shape. 
We speculate that this {\em pseudo-thermalization}, operating 
at the level of multiparton wave functions could perhaps explain
the apparently thermal  spectra observed in those high-energy processes,
where the conditions for a genuine thermalization are not met.
\par
     
\smallskip\noindent{LPT Orsay 02/31 (April 2002)}
\end{abstract}
\maketitle

Consider a multiparticle state in a finite box. If
the state were belonging to a classical statistical
ensemble, it would thermalize in the presence of an 
interaction between particles. In particular, the one-particle
momentum distribution would tend to the Boltzmann
form with time going on. Suppose now that the state is a 
pure quantum state. We wish to address the following 
questions: is it likely that the one-particle spectrum is 
approaching, for time $\to + \infty$, a limit independent of the 
initial state of the system, like in a generic Markov process? Is 
it plausible that this limit distribution looks thermal, reflecting 
the features of the corresponding classical ensemble? Although
these questions may sound academic, they are in fact motivated
by phenomenology. However, since the answers are, we believe,
of intrinsic theoretical interest, we postpone the discussion of
the phenomenological matters to the final paragraphs and
focus first on what appears to be, hopefully, more than an 
amusing speculation.
\par
It is difficult to approach the problem in its full
generality. Consequently, we adopt the
one-particle approximation: we follow a 
single
test particle moving in an external mean 
field, whose action represents the multiparticle 
environement. We know, that in a classical gas this motion 
generically leads to thermalization. A time arrow 
is attached to the motion of the test particle. This 
time arrow is also present when one goes over to 
the quantum description. This means that the evolution
generator describing the motion of the test particle is
non-hermitian. Hence, it will, in general, have
complex eigenvalues. Consequently, a single
quantum state of the test particle, independent of 
its initial state, will be selected for time 
$\to + \infty$. This state, to be regarded as the
stationary one, is reached by a process, which 
corresponds to the decay of a quantum fluctuation
and which is likely to resemble thermalization.
That is the general idea of this paper.
It will be illustrated by a simple example.
\par 
Consider a gas of classical particles first.
How does the thermalization occur? A given particle of the gas
is slowed down (accelerated)  when it collides with
a particle less (more) energetic than itself.
Eventually, its momentum fluctuates near the
average momentum measured by the gas
temperature. To mock this behavior assume
that our test particle gets random kicks, such that 
the transition $p \to p  \pm \delta$ occurs with
probability
\begin{equation}
Prob(p \to p  \pm \delta) \sim 
e^{\Delta^\pm(p)}
\label{2}
\end{equation}
where
\begin{equation}
\Delta^\pm(p) = \frac{1}{2T}[E(p) - E(p\pm\delta)]
\label{2bis}
\end{equation}
Here $E(p)$ is the energy of a particle. 
The transition probability (\ref{2})
satisfies the detailed balance equation
\begin{equation}
Prob(p) Prob(p  \to p  \pm \delta) = 
Prob(p  \pm \delta)Prob(p  \pm \delta \to p)
\label{3}
\end{equation}
with $Prob(p) \sim \exp{[-E(p)/T]}$, which is therefore 
the limit distribution of the process defined by 
(\ref{2}) \cite{foot3}.   
Equation (\ref{2})
violates the time reversal. It must be so, the process has the 
time arrow built into it, since it mocks the effect of 
the interaction of the test particle with its multiparticle
environement. 
\par
In order to simplify writing we assume  first that 
$\delta$ is a constant vector. Thus the motion is
one-dimensional. The extension of the discussion to 
a spectrum of momentum transfers is trivial.
\par
Eq. (\ref{2}) implies the following master equation for the
momentum probability distribution $w(p,t)$:
\begin{eqnarray}
\partial w(p, t) /\partial t =  
w(p + \delta, t) e^{\Delta^-(p + \delta)} \nonumber  \\
+ \; w(p - \delta, t) e^{\Delta^+(p - \delta)}
- w(p, t) [e^{\Delta^-(p)} + e^{\Delta^+(p)}]    
\label{fp}
\end{eqnarray}
where the time $t$ is measured in appropriate 
units, so that the interaction strength does not appear 
explicitly. It can be immediately seen that 
$w(p) = e^{- E(p)/T}$ is a stationary 
solution of (\ref{fp}), as expected. 
\par
We now go over to the quantum world and replace the 
particle by a wave. We assume that the kicks are due to the interaction
\begin{equation}
V =  \sum_p [ \eta^- e^{\Delta^-(p+\delta)} a^\dagger_p a_{p+\delta}
+  \eta^+ e^{\Delta^+(p-\delta)} a^\dagger_p a_{p-\delta}]
\label{8}
\end{equation}
where $\eta^\pm$ are some  phase factors, scalar functions 
of $p$ and $\delta$, while $a^\dagger_p$ ($a_p$) are 
time independent creation (annihilation) operators.
The interaction operator is not hermitian, the time 
reversal invariance is broken like in the classical model. 
\par
The number of particles is - by assumption - conserved 
\cite{foot4}. Hence the state at time $t$ has necessarily the form
\begin{equation}
\mid t \rangle = \sum_p \psi (p, t) a^\dagger_p \mid 0 \rangle
\label{9}
\end{equation}
The norm of this state is time dependent, the evolution being 
non-unitary. This does not matter for our purposes, since we are 
interested in this paper in the shape of the momentum distribution 
only. The Schroedinger equation describing the evolution 
of the wave packet (\ref{9}) reads
\begin{eqnarray}
i\partial \psi (p,t)/\partial t = 
\eta^- e^{\Delta^-(p+\delta)}\psi (p+\delta, t) \nonumber \\ 
+ \; \eta^+ e^{\Delta^+(p-\delta)} \psi (p-\delta, t)  
\label{10}
\end{eqnarray}
In order to determine the behavior of the solution we have 
to solve first the eigenvalue equation:
\begin{equation}
 \eta^- e^{\Delta^-(p+\delta)} \psi (p+\delta) +  
\eta^+ e^{\Delta^+(p-\delta)} \psi (p-\delta) = \lambda \psi (p)
\label{11}
\end{equation}
The transition probabilities are determined by the values of 
$\Delta^\pm$. In the absence of a better insight into the problem 
it is reasonable to assume that $\eta^\pm$ also depends primarily 
on this parameter. In order to obtain an analytically solvable model
we postulate a linear behavior 
\begin{equation}
 \mbox{\rm Arg}(\eta^\pm) = \xi_0 + \xi_1 \Delta^\pm (p\mp \delta)
\label{11bis}
\end{equation}
Set 
$\psi(p) =  e^{-E(p)(1+2i\xi_1)/2T} z(p)$  to get
\begin{equation}
 e^{i\xi_0}[z(p +\delta) + z(p-\delta)] = \lambda z(p)
\label{12}
\end{equation}
From here on the calculation becomes elementary. Assuming
periodic boundary conditions one has
\begin{equation}
 2 e^{i\xi_0}\;\cos{(x\cdot\delta)}
\;  \tilde{z}(x) =  \lambda \tilde{z}(x)
\label{13}
\end{equation}
where
\begin{equation}
 \tilde{z}(x) = \sum_p z(p) e^{ip\cdot x}
\label{14}
\end{equation}
Hence the sought eigenvalues and eigenvectors  are
\begin{equation}
\lambda_x = 2 e^{i\xi_0}\; \cos{(x\cdot\delta)}
\label{15}
\end{equation}
 and 
\begin{eqnarray}
\psi^{(s)}_x(p) = e^{-E(p)(1+2i\xi_1)/2T} \cos{(p\cdot x)}  \\ 
\psi^{(a)}_x(p) = e^{-E(p)(1+2i\xi_1)/2T} \sin{(p\cdot x)} 
\label{16}
\end{eqnarray}
When one considers a spectrum of momentum transfers, one gets
instead of (\ref{15}):
\begin{equation}
\lambda_x =  e^{i\xi_0} \; 
\sum_\delta \mu(\delta) \cos{(x\cdot\delta)} 
\label{15bis}
\end{equation}
with some $\mu(\delta)$, which we assume to be a positive weight
so that, generically, $x=0$ corresponds to the eigenvalue with
the largest modulus. The general solution of the Schroedinger equation is
\begin{equation}
\psi(p,t) = \sum_x [A_x \psi^{(a)}_x(p) + 
S_x \psi^{(s)}_x(p)] e^{-it\lambda_x}
\label{17}
\end{equation}
Let $0 < \xi_0 < \pi$.
Then only the contribution corresponding to $x=0$ 
survives in the limit $t \to + \infty$, because this is the one
whose norm is "explosing" the most rapidly (remember that the
state is not normalized; the $x=0$ eigenstate should be
regarded as the stationary solution). Consequently, the 
asymptotic spectrum is $w(p) \sim e^{-E(p)/T}$. This is the 
generic behavior, which can be avoided   by choosing very 
unnatural initial conditions only. At this point the goal we 
have set for ourselves at the beginning of this paper has been 
achieved. A rather robust quantum model has been constructed
illustrating the following conjecture: the evolution of the wave 
function of a system of very many interacting particles is likely 
to drive the one-particle spectrum towards a "thermal" shape. 
This behavior, which we would like to call 
{\em pseudo-thermalization}, is likely but not mandatory, 
in the model it occurs for a (wide) range of model parameters 
only. 
\par
Notice, that the physical contents of the classical and of
the quantum  model are    different. In the classical model 
a particle keeps changing its momentum. In the quantum 
model the components of a wave keep splitting. All
these components are a priori present at any time, only
their amplitudes get modified. However, in both models 
 the observable one-particle spectrum
has the same "thermal" shape. Moreover,
both models have the Markov property: the generic
limiting behavior is independent of the initial
state of the system.
\par
As long as one stays within the one-particle framework,
several specific assumptions of our quantum 
model can be relaxed. Still, even in this framework,
it would be interesting to check how generic is the
result obtained in this paper, viz.  does it hold for
processes which are significantly different from (\ref{2}).
The difficulty is mainly technical, at least if one looks
for an analytic solution.  However, going beyond the 
one-particle picture and checking the conjecture in a 
genuine multiparticle context requires a significant 
conceptual advance. On the theory side, we made just 
a first step and much remains to be done.
\par
As mentioned at the beginning of this text, our research
has actually a phenomenological motivation. Indeed, 
in high energy processes involving hadrons the inclusive 
spectra often look thermal, in circumstances where genuine
thermalization seems very unlikely. For example, in soft
hadronic collisions the inclusive transverse 
momentum spectrum at not too large  $p_\perp$ 
has an approximately boltzmannian shape, 
\begin{equation}
\frac{d\sigma}{d^2p_{\perp}} \sim e^{-E(p_\perp)/T} ,
\label{0}
\end{equation}
as if it resulted from a thermalization process occuring 
during the collision. Here, $E(p_\perp)$  denotes the 
"transverse energy" of a  secondary \cite{foot1} 
and $T$ is a "temperature"  parameter,
roughly independent of the secondary hadron species. 
However, the thermalization in the final
state, debatable but plausible in a heavy-ion collision, is
nearly impossible in a minimum bias nucleon-nucleon collision, 
where (\ref{0}) is the best observed. 
\par
As is well known, in the infinite momentum frame $P \to \infty$ 
a hadron can be regarded as a 2D nonrelativistic, coherent 
multiparton state enclosed in a box, with dimensions fixed by 
the confinement scale. The kinetic energy of a parton is 
$p_{\perp}^2/2xP$,  $x$ denoting its longitudinal 
momentum fraction. It is plausible that pseudo-thermalization
holds for the multiparton wave function. If so, then 
one can argue, invoking the parton-hadron duality,  
that after the hadron is broken by the interaction
the inclusive spectrum reflects the "thermal" 
distribution holding at the parton level \cite{foot2}. 
\par
Notice, that since $xP$ plays here the role of "mass",
one expects that the average transverse momentum is
larger for larger longitudinal momentum (in cms). Such
a correlation is indeed observed. On the other hand it is
not expected when the Schwinger mechanism is operating.
\par
Another example: in ref. \cite{bhal} a surprisingly
good and economic description of structure
functions is presented, using a statistical ansatz. This 
might be another example of pseudo-thermalization,
this time oberved directly at the parton level.
\par
On the phenomenology side, our exercise 
supports the idea that the "thermal" spectra
observed in high-energy processes involving hadrons
do actually reflect pseudo-thermalization 
affecting the structure of the
multiparton wave functions of the
incoming hadrons. If true, this is to large extent
independent of the details of the
microscopic dynamics - a common
feature of many multiparticle production
phenomena - except that the validity of
the parton picture requires an
asymptotically free theory, of course \cite{foot5}.
\par
{\bf Acknowledgements}: I am indebted to Roger Balian,
Andrzej Bialas, Ian Kogan and Jean-Pierre Leroy for helpful comments
and suggestions. This work was partially supported by
EC IHP grant HPRN-CT-1999-000161. Laboratoire de
Physique Th\'eorique is Unit\'e Mixte du CNRS UMR 8627.
\par
{\bf Note added}: Preparing the original version of this note I have
completedy overlooked results obtained in the context of so-called
quantum chaos theory, and in particular the early paper by M. Srednicki \cite{sred},
much more profound that the heuristic discussion presented above. Working with a 
gas of hard spheres, he has given strong arguments  in favor of the conjecture
that one-particle distribution in a stationary pure quantum state is
Boltzmann, provided the corresponding classical state is chaotic.


\begin{thebibliography}{99}
\bibitem{foot3}  We do not claim that the transition
probabilities are exactly given by (\ref{2}) in real life.
There is an infinity of stochastic processes yielding the
same limit distribution.
\bibitem{foot4} Particle number non-conservation
and quantum statistics effects are neglected in the
simple framework we have adopted.
\bibitem{foot1} The data
are best fitted with $E(p_\perp)$ set equal to the
transverse mass $\sqrt{p_{\perp}^2 +m^2}$. From
the theory point of view, one would perhaps prefer the
non-relativistic expression $p_\perp^2/2m$, because
the shape of the spectrum is expected to reflect the
galilean invariance of the original parton state. The
two choices are equivalent for soft enough $p_\perp$.
There is plenty of data the reader can consult.
Relatively recent ones can be found, for example,
in \cite{data}.
\bibitem{data} M. Aguilar-Benitez et al., Z. Phys. C50 (1991) 405.
\bibitem{foot2} In an alternative non-thermal
interpretation of (\ref{0}) one refers to the Schwinger mechanism
\cite{schw} of pair production in a uniform (chromo)electric
field \cite{strings}.
\bibitem{schw} J. Schwinger, Phys. Rev. 82 (1951) 664.
\bibitem{strings} E. Brezin and C. Itzykson, Phys. Rev. D2 (1970) 1191;\\
A. Casher, H. Neuberger and S. Nussinov, Phys. Rev. D20 (1979) 179;\\
B. Andersson, G. Gustafson and T. Sjoestrand, Z. Phys. C6 (1980) 235;\\
A. Bialas, Phys. Lett. B466 (1999) 301 .
\bibitem{bhal}R.S. Bhalerao, Phys.Lett. B380 (1996) 1;
R.S. Bhalerao, N.G. Kelkar, B. Ram, Phys.Lett. B476 (2000) 285;
R.S. Bhalerao, Phys.Rev. C63 (2001) 025208.
\bibitem{foot5} Transverse momentum distributions
of the form (\ref{0}) are also observed in
jets produced in $e^+e^-$ annihilations.
In principle, we do not see any major
obstacle in extending the speculations of
this paper to the evolution of the
chromodynamical state produced there.
\bibitem{sred} M. Srednicki, Phys. Rev. E50, (1994) 888. 
\end{thebibliography}
\end{document}